\documentstyle[aasms4]{article}
\begin{document}
\title{Jeans instability of interstellar gas
clouds in the background of weakly interacting massive particles}
\author{David Tsiklauri \footnote{email: tsiklauri@physci.uct.ac.za,
http://pc021.phy.uct.ac.za/tsiklauri/}}
\affil{Physics Department, University of Cape Town, Rondebosch 7701,
South Africa}
\begin{abstract}
Criterion of the Jeans instability of interstellar gas clouds which are
gravitationally coupled with weakly interacting massive particles
is revisited. It is established that presence of the dark matter
always reduces the Jeans length, and in turn, Jeans mass of the
interstellar gas clouds. Astrophysical implications of this
effect are discussed.
\end{abstract}

Investigation of physical processes in clouds of interstellar gas
is important for a number of reasons. There is a strong evidence that
the formation of stars occurs through gravitational collapse of
the interstellar gas clouds. An important mechanism in triggering
the formation of stars and/or stellar clusters is believed to be
a high-velocity (supersonic) cloud-cloud collisions 
in which a dense gaseous slab is formed with two plane-parallel
shock fronts propagating away from the interface (Usami et al.
1995). Then the slab grows in mass becoming unstable
against gravitational instability which causes its fragmentation.
The fragments or the gaseous clumps, in turn, collapse further
and evolve into stars and/or stellar clusters.
Yet another mechanism for triggering of star formation is
Rayleigh-Parker instability which leads to an increase in the
curvature of the magnetic field lines that initially were
directed parallel to the galactic plane. The interstellar
gas clouds move along the field lines (due to the "frozen in"
condition) and fall into the "wells" of the filed. This
causes further stretching of the magnetic field lines,
making "wells" deeper, thus large amounts of gas can
fall into the "wells" creating large concentrations of gas
which then becomes gravitationally unstable (Gorbatskii, 1977).

In this letter we consider stability of the interstellar gas clouds
against Jeans instability in the presence of background
weakly interacting massive particles (WIMP).
We find that presence of WIMP matter yields an unavoidable reduction of
Jeans length, Jeans mass and Jeans time (time-scale of the collapse
via Jeans instability).

Existence of WIMP matter, one of the
possible form of dark matter, the latter itself a
dominant mass component
of the universe, is strongly motivated both by
standard models of particle physics and cosmology.
Generally speaking,
mass density associated with the luminous matter
(stars, hydrogen clouds, x-ray gas in clusters, etc.)
cannot account for the
observed dynamics on galactic scales and above,  (Trimble,  1987),
thus, revealing the existence of
large amounts of dark matter or otherwise pointing to a
breakdown of Newtonian
dynamics (Milgrom, 1994, 1995) or the conventional law of
gravity (Mannheim, 1995).
The role of dark matter could be played by
anything from novel weakly interacting elementary
particles to normal matter
in some invisible form (brown dwarfs, primordial black holes,
cold molecular gas, etc.). A global, homogeneously
distributed dark matter component can be provided by the vacuum
in the form of a cosmological constant (see Raffelt, 1995 for a review).
Therefore, a revision of classical theory of Jeans instability in the
presence of WIMP matter, or generally speaking any type of
{\it microscopic}
dark matter which is coupled with an interstellar gas cloud only
via gravitational interaction, seems to be of a considerable
importance, especially in the context of star and/or stellar
cluster formation. 

The low-surface brightness dwarf spiral
galaxy DDO 154 has one of the most extended and best studied dark
matter halo rotation curves (Carignan \& Freeman, 1988;
Carignan \& Beaulieu, 1989; Burkert \&
Silk, 1997), with a precisely known contributions from
stars, gas and dark matter. 
DDO 154 is one of most gas-rich
systems known, but what is more important, the shape of its
rotation curve is completely dominated by the dark matter
even in the innermost regions. Thus, the galaxies like DDO 154
where gas dynamics is governed mostly by the dark matter
gravitational potential are the best examples where novel
effects revealed in this letter would be pronounced at most.
However, because of their relative proximity, detailed
observations of interstellar gas  clouds are possible
only for our galaxy and one has to extrapolate this
knowledge on other galaxies, because there is no reason
to believe that our own galaxy is exceptional in any
respect. Moreover, our own galaxy is known to have
a dark matter halo, which makes results presented here
relevant for the description of physical processes in the
interstellar gas clouds in the Milky Way (see discussion below).

Equations that govern dynamics of two self-gravitating
fluids (an interstellar gas cloud and WIMP matter) inter-coupled only
via gravitational interaction can be written in the following
way:
$$
{{\partial \rho_i}\over{\partial t}} + \nabla (\rho_i \vec V_i)=0,
\eqno(1)
$$
$$
{{\partial \vec V_i}\over{\partial t}} +
(\vec V_i \cdot \nabla) \vec V_i = - \nabla \varphi_i -
{{\nabla P_i}\over{\rho_i}}, \eqno(2)
$$
$$
P_i = K_i \rho_i^{\gamma_i}, \eqno(3)
$$
$$
\Delta \varphi_i = 4 \pi G \sum_i \rho_i, \eqno(4)
$$
where notation is standard: $\vec V$, $P_i$, $\rho_i$ and $\varphi_i$
denote velocity, pressure, mass density and gravitational potential
of the fluids. The subscript $i=g,w$ denotes two components,
interstellar gas and WIMP matter respectively.
Now, writing every physical quantity, for brevity commonly
denoted by $\vec f$, in a form of
$\vec f= \vec f^0 + \vec f^\prime \exp[{i( \vec k \vec r - \omega t)}]$
and doing usual linearization of the Eqs.(1)--(4)
we can obtain following dispersion relation for the perturbations
$$
\omega^2={{1}\over{2}}\left[ {
(\omega^J_g)^2 + (\omega^J_w)^2 \pm
\sqrt{\left[(\omega^J_g)^2 -(\omega^J_w)^2 \right]^2 + 4 \delta^2}
}\right], \eqno(5)
$$
where $(\omega^J_g)^2 \equiv 4 \pi G \rho^0_g - c_g^2 k^2$,
$(\omega^J_w)^2 \equiv 4 \pi G \rho^0_w - c_w^2 k^2$.
$c_g \equiv \sqrt{\gamma_g P^0_g/ \rho^0_g}$
and $c_w \equiv \sqrt{\gamma_w P^0_w/ \rho^0_w}$
denote speeds of sound of the gas and dark matter
respectively and finally,
$\delta \equiv 4 \pi G \sqrt{ \rho^0_g \rho^0_w}$,
which is the term providing the gravitational
coupling between the two fluids.

From Eq.(5) a criterion for the onset of instability
can be easily obtained
$$
(\omega^J_g)^2 (\omega^J_w)^2 < \delta^2.  \eqno(6)
$$
Now, introducing Jeans length for the gas and dark matter
in an usual way
$$
\lambda^J_g = c_g \sqrt{{\pi}\over{G \rho^0_g}} \;\;\;
\lambda^J_w = c_w \sqrt{{\pi}\over{G \rho^0_w}}, \eqno(7)
$$
we can rewrite the instability criterion, Eq.(6), as
$$
\lambda > {{\lambda^J_g}\over{\sqrt{1+ (\lambda^J_g /
\lambda^J_w)^2 }}} \equiv \lambda^J \eqno(8)
$$
Note that in the case of a gas cloud alone, i.e. without
the dark matter background, the criterion for the onset of
Jeans instability, Eq.(8), is simply $\lambda > \lambda^J_g$.
That is presence of the gravitationally coupled dark matter
component results in an additional factor of
$1 / \sqrt{1+ (\lambda^J_g / \lambda^J_w)^2 }$.
Thus, it is clear that in the case of
presence of the dark matter,
Jeans length for the gas cloud is always reduced to a
value smaller than without dark matter. The latter,
of course, is in accordance with general physical grounds,
since the dark matter induces additional gravitational pull
upon the gas inside the cloud. However, to the best of our
knowledge this effect has not been studied quantitatively
so far.

It is instructive to give a more specific estimate
for $\lambda^J$. For this purpose we assume, further, that a
proposed dark matter component is in a form of a neutrino
ball, one of the possible candidates of WIMP dark matter
which recently has been studied by some authors
(Bili\'c \& Viollier 1997; Bili\'c et al. 1997;
Tsiklauri \& Viollier, 1996, 1998a,b,c;
Viollier et al., 1992,1993; Viollier, 1994).
That is a gas cloud is immersed in a neutrino ball
which is composed of massive, self-gravitating, degenerate
neutrinos, an object in which self-gravity of the neutrino matter
is compensated by its degeneracy pressure, likewise in an ordinary
polytropic star except for thermal pressure is replaced by the degeneracy
pressure due to Pauli exclusion principle.
We start calculation with an estimate of the squared Jeans
length for WIMP matter, $(\lambda^J_w)^2$.
Using, Eq.(7) and  polytropic equation of state for the neutrino
matter (Viollier, 1994)
$$
P_\nu=K \rho^{5/3}_\nu, \;\;\; K =
\left({{6}\over{g}}\right)^{2/3}
{{\pi^{4/3} \hbar^2 }\over{5 m^{8/3}_\nu}} \eqno(9) 
$$
where, $g$ is spin degeneracy factor
(in these notations $g=1$ corresponds to the case when
there are only neutrinos in the ball, whereas $g=2$ corresponds
to the case when there are both neutrinos and antineutrinos
present),
$m_\nu$ is the neutrino mass, we obtain
$$
(\lambda^J_w)^2={\pi \over G} {5 \over 3} K \rho^{-1/3}. \eqno(10)
$$
Eqs.(1-8) have been derived under assumption that
the both gas and dark matter components are homogeneous.
In the neutrino ball, however, the physical quantities
vary with the radius, like in an ordinary
polytropic star. A gravitational potential of the
neutrino ball may well be approximated by a constant density
distribution, with an average density given by
$\bar \rho=(3 (-\xi^2 \theta^\prime)_1/ \xi^3_1) \rho_c$.
Here, $\rho_c$ denotes the central density in the neutrino ball
and $\xi_1=3.65375$ and $(-\xi^2 \theta^\prime)_1=2.71406$
are usual notations from polytropic theory of stars (Cox \& Giuli, 1968).
Thus, putting $(3 (-\xi^2 \theta^\prime)_1/ \xi^3_1)
\rho_c \to \rho$ in the Eq.(10) and using
standard definition of the Lane-Emden unit
of length (Cox \& Giuli, 1968) (except that we have $8 \pi$ instead of
$4 \pi$ since we have contribution both from neutrinos and
antineutrinos)
$$
r^2_n \equiv {5 \over 2} {1 \over {8 \pi G}} K \rho_c^{-1/3}, \eqno(11)
$$
we finally obtain
$$
(\lambda^J_w)= {{4 \pi}\over{3^{2/3}
(-\xi^2 \theta^\prime)_1^{1/6} \xi_1^{1/2}}} R_\nu =
2.6760 R_\nu. \eqno(12)
$$
Here, $R_\nu$ is the radius of a neutrino ball defined 
in an usual way by  $R_\nu=r_n \xi_1$.

Now, to estimate the reduction factor in the Jeans length, 
$1 / \sqrt{1+ (\lambda^J_g / \lambda^J_w)^2 }$,
we use a typical physical parameters for an interstellar
gas cloud, namely, $\lambda^J_g \approx 30$pc, its
size $\approx 2$ pc (Gorbatskii, 1986, 1977)
and assume that radius of a
neutrino ball is 1 pc, i.e. the gas cloud is fully immersed
in the neutrino ball. Using these values we obtain 
$1 / \sqrt{1+ (\lambda^J_g / \lambda^J_w)^2 }=8.8848 \times 10^{-2}$.
This estimate clearly demonstrates
that presence of background dark matter which is gravitationally
coupled to the gas cloud significantly
reduces its Jeans length, $\lambda^J_g$, and in turn,
its Jeans mass, which is proportional to $(\lambda^J_g)^3$.
Therefore we conclude that  the
star formation rate 
which is obviously related to the Jeans length
of a star-forming cloud can, in principle,  serve
as {\it a test for the amount of dark matter in the Galaxy}.

As it was mentioned above, typical physical parameters of an
interstellar gas cloud in our galaxy are a characteristic
size 2-3 pc, density $\approx 10^{-22}$ g$/$cm$^{-3}$, and
Jeans length $\lambda^J_g \approx$ 30 pc, i.e. a typical cloud is stable
against gravitational instability
(its size is an order of magnitude less than corresponding
Jeans length).
However,  as it is known
from the observations, in our galaxy there are
some gigantic cold gas clouds whose age is larger than 
their typical Jeans time. 
The lifetimes of these molecular clouds have been inferred to be
few times $10^7$ years from the total fraction of gas
mass in the Galaxy in the form of molecular gas,
and from the lifetimes of young stars associated with them 
(Blitz \& Shu, 1980). It has been argued that the shock
formation due to the observed  hypersonic velocities would
dissipate the energy stored in clouds quickly enough
so that entire cloud would collapse and form stars
in a time span not much longer than 
its free-fall time, that is,
$t_{\rm ff} =1.4 \times 10^6  /\sqrt{2 n (H_2)/ 10^3 cm^{-3}}$ yr,
where $n(H_2)$ is the number density of molecular hydrogen
(Goldreich \& Kwan, 1974; Field, 1978). Three approaches have
been put forward to explain this discrepancy (Mac Low, 1997). 
Arons \& Max (1975) 
suggested that strong enough magnetic fields
can prevent creation of shocks which will increase the
dissipation time. Scalo \& Pumphrey (1982) have argued
that hydrodynamic turbulence, which is thought to be capable
of preventing the collapse, will dissipate more slowly than
expected. However, as shown by Mac Low et al. (1997), 
via performed numerical studies of compressible, decaying
turbulence, with and without magnetic fields, the observed
long lifetimes and supersonic motions in molecular clouds
must be due to some kind of external driving
(e.g. stellar outflows (Silk \& Norman 1980), photoionization
(McKee 1989, Bertoldi \& McKee 1996), galactic shear (Fleck 1981)
of combination of these thereof) as undriven turbulence decays far too 
fast to account for observations.
On the other hand,
existence of the dark matter halo around our galaxy has been
thoroughly established through gravitational microlensing events
of stars in the Large Magellanic Could (Alcock et al., 1996, 1997).
It has been established that the dark matter within 50 kpc
(14 disk scale length) is order of $2.5 \times 10^{11} M_\odot$,
which is 4 to 5 times the mass of the galactic disk ($\approx 6 \times
10^{10} M_\odot$). Therefore, in the light of the results of this
letter, problem of existence of the  long-lived star-forming 
interstellar clouds of molecular gas,
pointed out above, becomes even more enigmatic.
At the same time, one has to admit that obviously, simple minded,
classical theory of Jeans instability cannot be used for
drawing of detailed picture of star formation process in the
molecular clouds. It is likely that mechanisms that trigger and
govern star formation are related to those, yet not clearly
identified, processes which prevent the gigantic molecular
clouds from gravitational collapse. However, it is probably
also not a mere coincidence that, for typical molecular
clouds having mass in the range $10^3$--$10^6 M_\odot$, 
number densities of the order of $10^6$ cm$^{-3}$ and 
temperature of the order of 10K, classical Jeans theory
sets correct mass scale, i.e. Jeans mass is of the order of 
$\sim  M_\odot$ --- a mass of a typical star.

Finally, we would like to remark that as an illustrative
example we have considered a neutrino ball as a possible
from of a dark matter. However, the results of this work
are equally valid for any type of microscopic dark matter,
microscopic in a sense that it can be described
as a continuous medium, i.e. by the hydrodynamic equations.
\vskip 1cm
I would like to thank Dr. Steven  Shore
of Indiana University South Bend for valuable comments and 
suggestions. I also would like to thank
an anonymous referee for a number of useful suggestions.

\end{document}